\documentclass[fleqn,10pt]{wlscirep}
\usepackage[utf8]{inputenc}
\usepackage[T1]{fontenc}
\usepackage{ulem}

\usepackage{xcolor}
\usepackage[separate-uncertainty=true]{siunitx} 
\DeclareSIUnit\Molar{\textsc{m}}
\newcommand{\RE}[1]{\textcolor{black}{#1}}

\title{Color-switching hydrogels as integrated microfluidic pressure sensors}

\author[1,+]{Lucie Duclou\'e}
\author[2,4, +]{Md. Anamul Haque}
\author[1]{Martyna Goral}
\author[2]{Muhammad Ilyas}
\author[2,3]{Jian Ping Gong}
\author[1,*]{Anke Lindner}
\affil[1]{Laboratoire de Physique et M\'ecanique des Milieux H\'et\'erog\`enes, UMR 7636, CNRS, ESPCI Paris, PSL Research University, Universit\'e Paris Cité, Sorbonne Universit\'e, Paris, FR-75005, France}
\affil[2]{Laboratory of Soft \& Wet Matter, Faculty of Advanced Life Science, Hokkaido University, Sapporo, Hokkaido 001-0021, Japan}
\affil[3]{Institute for Chemical Reaction Design and Discovery (WPI-ICReDD), Hokkaido University, Sapporo 001-0021, Japan
}
\affil[4]{Department of Chemistry, University of Dhaka, Dhaka-1000, Bangladesh}

\affil[*]{anke.lindner@espci.psl.eu}

\affil[+]{these authors contributed equally to this work}

\begin{abstract}
Precisely measuring pressure in microfluidic flows is essential for flow control, fluid characterization, and monitoring, but faces specific challenges such as \RE{achieving} sufficient resolution, non-invasiveness, or ease of use. Here, we demonstrate a fully integrated multiplexed optofluidic pressure sensor, entirely decoupled from the flow path, that enables local pressure measurements along any microfluidic channel without altering its flow geometry. The sensor itself relies on the compression of a soft mechano-actuated hydrogel, changing color in response to a pressure change. The hydrogel is separated from the fluid circulating in the channel by a thin membrane, allowing for the unrestricted use of different types of fluids.  Imaging the gel through the transparent PDMS with a color camera provides a direct, easy, and contact-free determination of the fluid pressure at the sensing location for pressures as small as \SI{20}{\milli\bar} with a resolution of around \SI{10}{\milli\bar}. The sensitivity and accessible pressure range can be tuned via the mechanical properties \RE{of the sensing unit}. The photonic gel can also be used to acquire 2D pressure or deformation maps, taking advantage of the fast response time and fine spatial resolution. 
\end{abstract}
\begin{document}

\flushbottom
\maketitle
%
%
\thispagestyle{empty}

\section*{Introduction}


The widening applications of microfluidic devices in fields as different as biomedicine, organ-on-chip, chemical analysis, material characterization or water purification, have increased the need for reliable pressure-sensing in those systems. pressure-sensing in such small fluidic circuits, however, faces specific challenges stemming from the design of microfluidic chips. The lab-on-a-chip layout requires that the sensor be fully integrable to the chip, and so most commonly to PDMS architectures.  The sensor sensitivity is challenging, given that the pressure changes along short fluid sections can be minute. In addition, the pressure sensor is required to be non-invasive and, as such, not to modify the local flow geometry.  The sensor is preferably easy to use and read out.  \\

Several methods have been proposed to overcome these difficulties, which rely on soft-sensing elements that deform under pressure. This deformation is usually measured via electric methods, which can be piezoelectric pillars~\cite{LiMicroelEng2010}, conductive liquid circuits~\cite{ParkJMM2010}, conductive inflating membranes~\cite{WangBiomicroflu2009} or capacitive sensors based on composite foam materials \cite{Pruvost2019,Gauthier2021}. Such methods require that the microfluidic chip is wired to an external electrical apparatus, which may not be convenient. Optical methods for the detection of the deformation of sensing elements have been proposed, and overcome this drawback. The deformation of an optical grating~\cite{KazuoJMM2002}, interference patterns in a deformed cavity~\cite{SongOptLett2010}, light focusing through an inflating membrane~\cite{OrthLoC2011}, color intensity change in a concentric chamber \cite{Tsai2016}, the displacement of a colored liquid or a liquid-gas interface \cite{Zhang2021, Araci2014} and very recently, nano-structured photonic inflating membranes~\cite{EscuderoSciRep2019} have been used.  \RE{All those methods enable remote sensing of the fluid pressure, through precise calibration of the optical detection of the small deformations involved. However,} the pressure measurements rely on the deformation of a sensing element and thus modify the microfluidic flow geometry. The pressure drop in a flat channel scales with the square of the channel height and such a deformation can thus have important consequences on fluid flow and strongly modify the local pressure. The work by Orth \textit{et al.}~\cite{OrthLoC2011} stands out as a realization in which the sensing element is decoupled from the flow path of the fluid: therefore, the deformation of the sensor does not modify the geometry of the channel in which the fluid flows. \\

Here we propose a different sensing method, directly integrating a material that changes color under applied pressure into a microchannel (see \cite{Ducloue2023}). Such mechano-chromic materials are for example pressure sensitive paint \cite{Huang2015}, colorimetric films taking advantage of the photonic shift during the disassembly of gold nanoparticle chains \cite{Han2014}, nanostructured photonic membranes \cite{EscuderoSciRep2019} or the photonic hydrogel used here \cite{Yue2014mechano}. The use of these materials remains scarce for optofluidic applications, with the exception of the work mentioned above by Escudero \textit{ et al.} \cite{EscuderoSciRep2019} where a color shift is elegantly used to detect membrane deflection or the integration of pressure-sensitive paint, only suitable for gas flows and requiring observation in a dark environment using UV light. Contrary to those applications, we directly use the color change in visible light of a novel soft photonic hydrogel~\cite{Yue2014mechano}. The gel is actuated under very small compressive stress and exhibits an ultrafast response time and high spatial resolution \RE{(\SI{0.1}{ms} and \SI{10}{\um}-\SI{100}{\um} respectively)} \cite{Yue2014mechano}. We built a multilayered PDMS sensing unit composed of a pressure-sensing cavity separated from the mechanochromic gel by a thin membrane. Such sensing units can be fully integrated into microfluidic devices, by, for example, connecting them at desired locations to microchannels, leading to a noninvasive multiplexed pressure measurement. As the membrane separates the gel from the fluid circulating in the microchip, the sensor can be used with all types of liquid or air, while avoiding the presence of liquid/gas interfaces. We take advantage of the softness of the gel to demonstrate pressure measurements at very low pressures (from \SI{20}{\milli\bar} to \RE{\SI{140}{\milli\bar}}) with high \RE{resolution} of around \SI{10}{\milli\bar}. The combination between membrane \RE{bending} stiffness, cavity size, and \RE{theoretically gel properties (not systematically varied in this work),} allows for fine-tuning of the measurement range and in particular the extension of the pressure range towards higher pressures.  Additionally, the gel can also be used to obtain 2D pressure maps and to monitor spatial and temporal pressure differences or determine local deformations.

\section*{Results}
\subsection*{General design and operation principle}

\begin{figure}[h]
	\centering
	\includegraphics[width=1\linewidth]{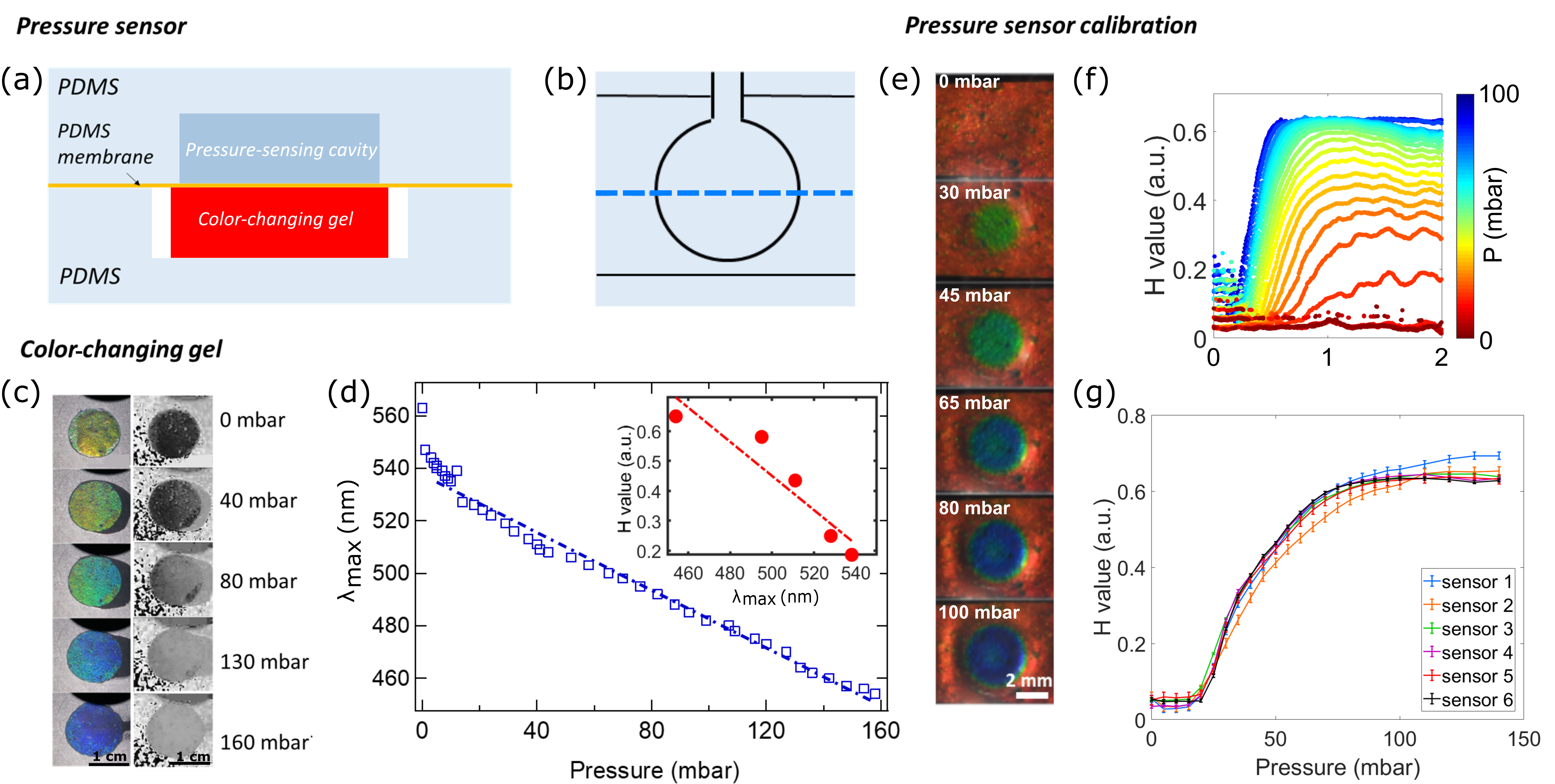}
	\caption{{\bf General design of the pressure-sensing unit} Schematic of the cross-sectional view (a) and the top view (b) of the assembled sensor (not scaled). The dotted line indicates the position of the cross-sectional view. {\bf Color-changing hydrogel} (c) Color-response of a sample of a free-standing soft photonic gel of \SI{1}{\cm} diameter under uniform applied pressure \RE{between a glass slide and a substrate (see Supplementary Fig. 1)}. The left row indicates the visual response of the gel with RGB color images and the right row with hue value intensity images. (d) Maximum reflected wavelength measured using a spectrometer as a function of applied pressure.  \RE{The line is a liner adjustment, $R^2=0.98$}. \RE{Inset: hue value as a function of the reflected wavelength. The line is a guide for the eye.} {\bf Calibration of the sensing unit.} (e) Top view of a sensor (diameter \SI{4}{\mm}, gel layer thickness \SI{140}{\um}, PDMS membrane thickness \SI{100}{\um}) connected to a pressure controller to increase applied static pressure (from \SI{0}{\milli\bar} to \RE{\SI{100}{\milli\bar}}). (f) \RE{Average} hue value profile across the \RE{radius} of the sensor, \RE{for pressures varying from \SI{0}{\milli\bar} to \SI{100}{\milli\bar} with an increment of \SI{5}{\milli\bar}.} (g) Calibration curve of several \RE{\SI{4}{\mm} diameter sensors placed above the same gel slab}: hue value as a function of applied pressure.}
 \label{fig:1}
\end{figure}

The pressure-sensing unit is designed as a multilayered PDMS system, illustrated schematically in Fig. \ref{fig:1}~(a,b).  The sensor is made of two molded PDMS layers, obtained by standard soft lithography techniques~\cite{Tabeling2005} and separated by a thin PDMS membrane. The top layer contains a pressure-sensing cavity that is filled with the pressurized fluid to be analyzed. The bottom layer contains a \RE{chamber} enclosing a slab of the photonic hydrogel. The hydrogel being immersed in an aqueous buffer solution, the gel \RE{chamber} has an inlet and an outlet to refresh the buffer \RE{and is slightly larger than the gel slab to enable its lateral expansion while being compressed}. Details on the assembly are given in the Materials and Methods section. Special care is taken to match the height of the hydrogel slab and the \RE{chamber}, to avoid pre-compression of the gel or an offset in the pressure measurements. For convenience the upper cavity has a disc-like shape but is not limited to this choice. The pressure-sensing cavity fully covers the hydrogel slab as can be seen in Fig. \ref{fig:1}~(a,b). Under increasing pressure in the pressure-sensing cavity, the membrane deforms, and the underlying gel is compressed, subsequently changing color. In our setup, this color change is observed in reflection from above via a camera. The quality of the color observation is improved when the sensing unit is placed on a black background.

The soft photonic hydrogel has already been described before~\cite{Yue2014mechano}, and the details of its fabrication and characterization can be found in the Materials and Methods section. The layered gel has a lamellar structure formed by thin, rigid, and light-reflecting bilayers embedded in a soft hydrogel matrix. Typical interlayer distances are comparable to the wavelength of visible light. As a consequence, the multilayer structure selectively diffracts visible light, so that it appears colored under reflective white illumination. Under compression, the layer spacing decreases leading to a color change of the gel. The gel diffracts longer wavelengths in its undeformed state (\RE{orange} color), and shorter wavelengths when compressed (gradual shift to blue color). Typical gel thicknesses are between \SI{100}{\um} and \SI{1}{\mm} with \RE{Young's} moduli between \SI{10}{\kilo\pascal} and \SI{100}{\kilo\pascal} (see Supplemental Fig.~2(b)). Thinner gels can be fabricated, but the quality of the color of the reflected light is found to be reduced and in this work no gels below \SI{100}{\um} have been used.
\RE{A color change from orange (wavelength $\lambda\approx$ \SI{600}{\nm}) to blue (wavelength $\lambda\approx$ \SI{450}{\nm}) corresponds to a decrease in layer spacing and thus a compressive strain of the gel of $\epsilon=\Delta \lambda_{max}/\lambda\approx 0.25$.} For the thin gel this corresponds to a decrease in height of about \SI{25}{\um} and for the thick gel of \SI{250}{\um}. To be more compatible with the dimensions of a microfluidic chip, we chose to mainly work with thin gels and present the following results for a gel of thickness \SI{140}{\um}, but successful tests with thick gels have also been performed.

Snapshots of a free-standing gel sample (thickness \SI{140}{\um} at rest), \RE{not yet inserted into the microfluidic pressure-sensing unit},are shown in Fig. \ref{fig:1}~(c). Under the application of increasing uniform pressure, the color change from orange/yellow towards blue is visible on the left panel. The precise color response is measured with a spectrometer for each pressure as the maximum of a sharply-peaked distribution of the reflected wavelength~\cite{Haque2010}. \RE{Details about the setup and the measured color spectra are given in the Materials and Methods section and Supplemental Materials.}  The color response curve obtained with the spectrometer is given in Fig. \ref{fig:1}~(d) in terms of the maximum reflected wavelength $\lambda_{max}$ as a function of the applied pressure $P$. The decrease in wavelength (corresponding to a compression of the gel) \RE{\cite{Yue2014mechano, Haque2010}} is observed to be linear for most of the range of pressures tested. A stiffening of the gel for even larger deformations (toward a color response in violet) has been observed, but this is not shown here. The \RE{Young's} modulus of the gel sample tested has been determined to be approximately $Y$=\SI{90}{\kilo\pascal} \RE{(Supplemental Fig. 2~(b))}. Note that small changes in the fabrication process and gel swelling can lead to variations in the gel properties, such as initial layer spacing or gel modulus. For instance for the gel shown in Fig. \ref{fig:1}~(c) the color at rest was not perfectly \RE{orange}, but rather yellow preventing the full range of the visible light to be used. \RE{We perform a calibration step for each realization of the pressure-sensing unit to account for the precise properties of the gel sample used (see below)}.  \\

 For general pressure-sensing applications, using a spectrometer is not necessarily convenient, and it can be useful to measure and quantify the color switching of the gel with simpler equipment, such as a color camera or an optical fiber. This can be done by transforming color pictures as those shown in Fig. \ref{fig:1}~(c) left row into a gray-scale intensity map of the hue value $H$ (Hue-Saturation-Value color model,  obtained by image treatment) and shown in the right row. The diffracted light being monochromatic, the $H$ value derived from camera images is a good proxy for the wavelength. \RE{This is shown in the inset of Fig. \ref{fig:1}~(d), which compares the $H$ value calculated from images acquired with a color camera and the maximum reflected wavelength measured with a spectrometer. In the range between \SI{480}{nm} and \SI{540}{nm} $H$ decreases with $\lambda_{max}$, and can thus be used to quantify the color response of the gel simply from color images acquired with a camera, as done in this work.  A saturation of the hue value is observed for wavelengths below \SI{480}{nm}.}

\subsection*{Calibration of the pressure-sensing unit}

pressure-sensing units need to be calibrated to enable quantitative pressure measurements. The calibration step is required to quantify slight changes in the gel properties depending on the fabrication, hydrolyzation, and swelling of the latter, but also to account for the possible coupling between the membrane and the gel or the influence of the soft PDMS foundation on gel deformation. All sensors used here have a cavity height of \SI{100}{\um}, a PDMS membrane thickness of \SI{100}{\um}, and a gel thickness of \SI{140}{\um}. Two different sensing cavity sizes were used, of diameter \SI{4}{\mm} and \SI{1}{\mm} respectively, enabling different ranges of measured pressure. 
In Fig. \ref{fig:1}~(e) we show the calibration step for sensing units with the large cavity diameter of \SI{4}{\mm}, sensitive to smaller pressures compared to the \SI{1}{\mm} diameter sensor (see Supplemental Fig. 2~(c)). Several cavities have been placed on top of a long slab of hydrogel and connected to a microchannel for simple pressure application.  The calibration step is performed by filling the sensing cavities with water and applying static pressure with the help of a pressure controller (Fluigent Lineup Flow EZ). One can rely either on the permeability of PDMS to evacuate the air contained in the sensing cavities or connect the sensing cavities to an outlet to directly evacuate the trapped air. \RE{The outlets are closed during the calibration step so that the static pressure can be applied to the system without flow.} The pressure is increased stepwise and a color image of each sensor is taken for each pressure value. Examples of pictures of the color evolution during this procedure are given in Fig. \ref{fig:1}~(e). At zero applied pressure, the gel strip under the sensors is uniform orange in color. When the pressure is increased, the sensing cavity bottom membrane presses on the gel, making it locally compressed: circular spots appear on the gel strip, which shift toward blue colors as the pressure is increased. \\

The color profile for each sensor is then analyzed in terms of the $H$ value. Due to the circular shape of the sensors, this profile is axisymmetric. To get optimal resolution, we detect the center of each sensor and average the $H$ value over the azimuthal direction.  The resulting radial $H$ profile is illustrated in Fig. \ref{fig:1}~(f) for a given sensor. As the reflected wavelength decreases linearly with the vertical compression of the gel, the profile of the $H$ value is also a proxy for the deformation of the gel below the membrane. For all $H$ value profiles one observes no deformation at the cavity borders where the membrane is attached to the PDMS layer. The deformation increases towards the middle of the cavity, where a plateau is reached for smaller pressures (\SI{20}{\milli\bar}-\SI{40}{\milli\bar}) and thus for small deformations. For larger deformations (\SI{40}{\milli\bar}-\SI{80}{\milli\bar}) more complex profiles are observed, with a small dip in deformation in the middle of the cavity. For the largest pressures used, above \SI{80}{\milli\bar}, the sensor saturates and the profile becomes completely flat. \RE{This saturation can result from two origins, first the saturation of the hue value for small wavelengths and second the stiffening of the thin gel layer under compression. }

Due to the complex shape of the $H$ value profile at larger pressures, we use the value of $H$ at $R/2$, where $R$ is the radius of the sensors, to build the calibration curves. Fig. \ref{fig:1}~(g) shows such curves for \RE{the 6 tested pressure sensors positioned above the same gel slab}. For pressures below \SI{20}{\milli\bar}, no color change is observed. We suspect that this is due to a small gap between the gel strip and the membrane, so the small deformation of the membrane under low pressures is not sufficient to establish contact with the gel.  At the highest pressures tested, the sensors saturate. \RE{For optimal precision, the exact shape of the curve is used for calibration purposes for intermediate values.} The calibration captures, as already pointed out, \RE{any variations that might occur in gel properties from one batch to another}, the characteristics of the sensor geometry and the specific features of the setup used, such as the illumination conditions. Within one experimental realization, very good agreement between the curves is obtained ( Fig. \ref{fig:1}~(f))  confirming the uniformity of the gel and the good repeatability within one set of sensors.The working range of the sensors with diameter \SI{4}{\mm} presented here is found to be from \SI{20}{\milli\bar} to \SI{80}{\milli\bar}, and for the \SI{1}{\mm} diameter sensor (\RE{calibration Supplemental Fig. 2~(b)}) from \SI{60}{\milli\bar} to \SI{140}{\milli\bar}.

\subsection*{Membrane-gel coupling and working range of pressure-sensing cavities}

To better understand the mechanical coupling of the membrane with the color-changing gel, we have modeled the sensing unit using a thin-plate approximation for the membrane, described as a circular bending plate, clamped at the edges and resting on an elastic foundation ~\cite{timoshenko1959theory}. The coupled response of the membrane and gel under varying pressure and sensor radius is presented in the Supplementary Material. Qualitatively the shape of the predicted deformation profiles is in very good agreement with the experimental observations in Fig. \ref{fig:1}~(f), including the small dip that develops in the middle at strong compression.

The model also highlights how decreasing the radius of the cavity leads to less deformation for identical applied pressures and thus to an effective stiffening of the membrane/gel system. This principle can be used to extend the accessible pressure range to higher pressures and explains that in our system the smaller cavities react to a different pressure range as compared to the larger cavities.
For an increasingly large cavity radius, we expect the role of the membrane to be less and less important and the deformation of the gel to be only given by the gel \RE{Young's} modulus, defining in this way the smallest pressures accessible.

Thus, the working range of the sensors can be tuned by playing with the mechanical coupling of the membrane to the gel. For large sensors, the role of the membrane can be neglected and the limiting factor is the gel \RE{Young's} modulus itself.  The smallest measurable pressure (as well as the resolution of the pressure measurement) is given by the modulus of the gel and the accuracy with which the color change can be detected. \RE{The calibration curves of the large sensors shown in Fig. \ref{fig:1}~(g) show hue values obtained at pressure increments of \SI{5}{\milli\bar}. taking also into account small variations between the different sensors we estimate the resolution of our sensors to be of \SI{10}{\milli\bar}. Due to the imperfect alignment of the membrane, the smallest value measured with our sensor is slightly larger and corresponds to \SI{20}{\milli\bar}}. Higher pressures can be reached when working with smaller cavities, where the membrane plays a stiffening role. Tuning more precisely the parameters of the model presented in Supplementary Materials could help the design of sensors with the desired pressure range in the future.

\section*{Validation of the sensor under flow}

\subsection*{Multiplexed integration into a microchannel}

To measure the pressure inside microfluidic channels of variable design, the sensing units can be directly integrated into microfluidic chips, and we demonstrate here a realization using a straight square channel. In order to avoid deformation of the flow channel geometry when the membrane deflects under pressure, we take inspiration from the design introduced by Orth \textit{et al.} ~\cite{OrthLoC2011}. We decouple the sensing area by connecting the sensing units laterally to the flow channel. The top view of our chip is shown in Fig. \ref{fig:2}~(a): the main flow channel is a straight channel of length $L=$\SI{20}{\mm} and quasi square cross-section of width $w = $\SI{110}{\um} and height $h = $\SI{100}{\um}. Equally spaced sensing units are integrated into the chip and connected via narrow channels (width \SI{50}{\um}) to the main channel. The sensing cavities are embedded in the same PDMS layer as the flow channel and, as previously separated from the hydrogel \RE{chamber} in the second PDMS layer by a thin membrane (see Fig. \ref{fig:1}a,b). Each sensor is connected to an outlet opposite to its side channel (see Materials and Methods). Two sets of sensors have been placed in this chip, with 6 sensors on each side of the main flow channel.  Each series of sensors rests on a single photonic gel slab, which appears in \RE{orange} in the picture (Fig. \ref{fig:2}b).  Outlets (flow channel and sensors) are left open during device filling and are closed for experiments \RE{of imposed static pressure} so that only the main flow channel remains connected to fluid sources during normal operation of the device, thus avoiding flow in the lateral channels and cavities. \RE{In the case of imposed flow in the channel, its outlet is left open, while the sensor outlets remain closed}. \\

\subsection*{Flow measurements}

\begin{figure}[ht]
\centering
\includegraphics[width=1\linewidth]{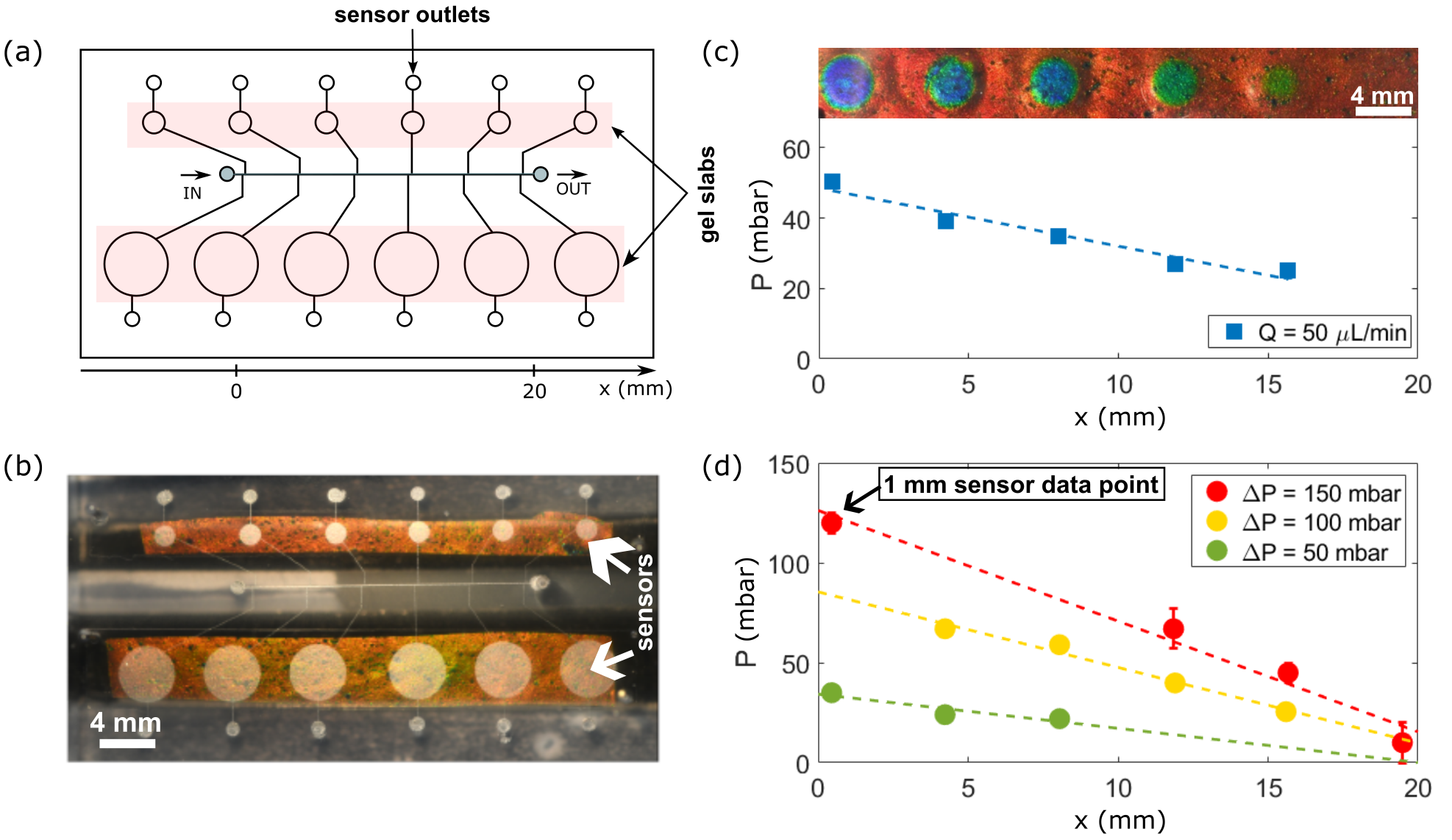}
\caption{{\bf Schematic of the channel design} (a) top view and (b) picture of the experimental realization, showing the orange color of the gel. \RE{The chip used is the one calibrated in Fig.~\ref{fig:1}(g) and Supplementay Fig. 2(c).} {\bf Response of the pressure sensors} (c) under steady water flow at \SI{50}{\micro\liter\per\min} (blue squares). Inset: top view of the sensors at a steady flow of \SI{50}{\micro\liter\per\min}. (d) Under imposed \RE{pressure difference} $\Delta P$ (bottom to top: \SI{50}{\milli\bar}, \SI{100}{\milli\bar}, \SI{150}{\milli\bar}). \RE{The response is measured from the \SI{4}{\mm} diameter sensors except for one data point (indicated in the figure) which is from a \SI{1}{\mm} diameter sensor. The pressure response as a function of distance is linear in all cases ($R^2$ values bottom to top: 0.86, 0.98, 0.98)}. Error bars are often smaller than the symbol size and thus not visible on the figure. The error corresponds to the standard deviation.}
\label{fig:2}
\end{figure}

Once the device has been calibrated, according to the procedure described above, it can be used for local pressure-sensing in flowing liquids. We first tested our pressure sensor by imposing a steady flow of water (flow rate: \SI{50}{\micro\liter\per\min}) in the straight channel and leaving the outlet open to the atmosphere \RE{while the sensor outlets remained closed}. The response of the large \SI{4}{\mm} sensors to this flow is shown in the inset of Fig. \ref{fig:2}~(c): a pressure gradient is visible along the channel. Note that the pressure at the last sensing location is not large enough to activate the last sensor: it is below \SI{20}{\milli\bar}. Using the calibration curves of Fig. \ref{fig:1}~(g), the pressure profile in the channel can be measured and is plotted with blue squares in Fig. \ref{fig:2}~(c). This profile is linear, as expected, and gives a total pressure drop of about \SI{25}{\milli\bar} over \SI{15}{\mm}, which is within the uncertainties of the imposed flow rate \RE{(estimated to be 10\%)} and the channel dimensions \RE{(estimated to be 5\%)} in reasonable agreement with the \SI{30\pm10}{\milli\bar} drop expected from the hydraulic resistance of the channel: $\Delta P = Q R_H$ with $R_H = \frac{12}{1-0.62 (h/w)} \eta L \frac{1}{w h^3}$ with $\eta$ the viscosity and $w,h$ the width and height of the channel, respectively \cite{bruus2007theoretical}.  Note that the pressure at the channel outlet is not necessarily exactly zero even though no tubing was connected at the outlet. This is due to the possibility of small liquid droplets accumulating at the outlet and the small but finite height of the water column in the outlet.

The linearity of the measured pressure gradient under steady flow has been checked over a wider range by imposing a constant over-pressure at the inlet (the outlet remaining at atmospheric pressure). 
Note that part of the applied pressure drop occurs at the tubing connecting the channel to the fluid reservoir, leading to a smaller pressure drop along the channel compared to the imposed pressure difference. In addition, small changes in the tubing might also slightly modify this pressure drop and thus lead to different offsets at the in and outlet for different measurements. The measured pressure response is shown in Fig. \ref{fig:2}~(d) for 3 imposed pressure drops $\Delta P$. All profiles are linear as expected and show an increasing slope with increasing imposed pressure. At small pressure drops, the sensors closer to the \RE{channel} outlet cannot be used (lowest $\Delta P$) as the expected pressure is below the minimal resolution of \SI{20}{\milli\bar}. For very large pressure drops, the large sensors saturate. The small sensors can however provide complementary information and were used for the highest data point \RE{for the pressure difference of} $\Delta P =$ \SI{150}{\milli\bar} \RE{as indicated by an arrow towards the data point of the figure. The calibration curve for the \SI{1}{mm} sensor is provided in the Supplementary Materials Fig.~2(c)}. This data set demonstrates that a larger pressure range can be obtained by combining different cavity sizes of the sensing units.   

In summary, the pressure-sensing unit integration into a microfluidic chip demonstrates their potential for noninvasive pressure measurement of hydrodynamic pressure underflow. All pressure profiles measured are linear as expected and show an increasing slope with increasing applied pressure at the channel inlet. For an imposed flow rate, the pressure drop measured over the length of the channel is quantitatively correlated with the calculated pressure drop. The pressure measurements obtained with sensors of different diameters fall onto a single curve and expand the range of accessible pressures.

\subsection*{Spatially resolved pressure mapping}

\begin{figure}[ht]
\centering
\includegraphics[width=1\linewidth]{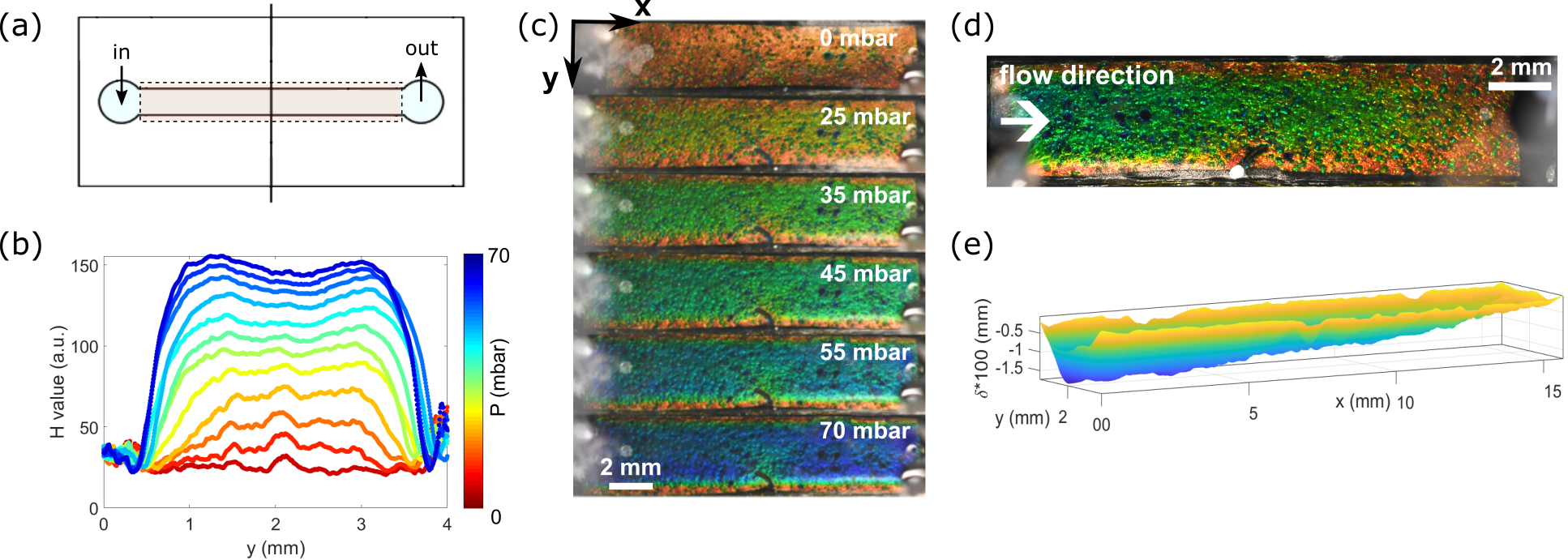}
\caption{
{\bf 2D pressure-sensing} (a) Schematic of the chip for 2D pressure-sensing. \RE{The line indicates the cross-section of hue value averaging. (b) Average measured hue value profiles from images presented in Fig. (c). (c) Top view of the chip: the rectangular flow channel has the same width and length as the overlaid gel channel. The uniform pressure loading increases from top to bottom.} \textbf{Non-uniform pressure as imposed by a steady flow of liquid :} (d) visual information and (e) reconstruction of the membrane deformation using the calibration obtained from images shown in Fig. (b).}
\label{fig:3}
\end{figure}

To demonstrate the good spatial resolution of our gel we have fabricated a device where a flow channel is directly positioned above a slab of color-changing gel, separated by the PDMS membrane. A schematic can be seen in Fig. \ref{fig:3}~(a). The change in color under static pressure is first tested and shown in Fig. \ref{fig:3}~(\RE{c}). The color change is smaller close to the channel walls where the membrane is fixed to the channel and increases towards the middle of the channel. Similarly to the pressure-sensing units (Fig. \ref{fig:1}) a calibration step is performed, this time using hue value profiles across the channel width (Fig. \ref{fig:3}~(\RE{b}). The maximum value in the center of the channel is used for calibration against the applied static pressure. 

When applying flow to the channel, the change in color along it is visible in the center of the channel (Fig. \ref{fig:3}~(d)) and can be used to infer the local pressure as a function of the position along the channel. Such spatially resolved local pressure measurements could be very useful for applications in more complex flow geometries such as branched channel networks, or channels with varying cross sections. A drawback of this implementation is however that, as discussed above, the compression of the gel leads to a significant change in local channel height and thus a modification of the flow profile and the hydrodynamic pressure. The pressure gradient in the straight channel shown here is thus not expected to be constant.

Instead of using the gel as a local pressure sensor, it can also be used as a local deformation sensor. As stated above, the color change of the gel observed under application of pressure results from a decrease in layer spacing of the lamellar gel and thus a change of the constructively reflected wavelength. Therefore, a color change is directly proportional to a decrease in layer spacing, and thus to an overall gel compression. The latter can be quantified by evaluating the change in the hue value and linking it (see the inset of Fig. \ref{fig:1}~(d)) to the change of the wavelength of the reflected light. From the change in wavelength, the change in layer spacing and thus the total strain of the gel slab can be evaluated. Knowing the thickness of the gel slab, we obtain the absolute local deformation. Fig. \ref{fig:3}~(e) shows the local deformation and therefore the local increase in channel height due to compression of the underlying gel. The $z$- axis has been multiplied by 100 to make the small deformations of at maximum slightly more than \SI{15}{\um} visible. The increase in deformation is clearly seen when going from the channel wall toward the channel center. Furthermore, the deformation at the center of the channel decreases continuously when going from the channel inlet to the channel outlet. Using the resolution of 3\% with which the change in wavelength can be measured, as estimated above, we expect to be able to resolve deformations with a resolution of around \SI{4}{\um}. This is in visual agreement with Fig. \ref{fig:3}~(e) indicating the very good resolution of our sensor. The spatial resolution in the $x$ and $y$ direction spanning the bottom surface of the channel can be visually estimated from the noise in the profiles of Fig. \ref{fig:3}~(\RE{b}) to be between \SI{50}{\um} and \SI{100}{\um} in agreement with Yue \textit{et al.} \cite{Yue2014mechano}. Our results have not been optimized \RE{to improve} the spatial resolution and the values given can be considered as an upper bound.

Deformation of PDMS channels under applied pressure is a common problem in these soft devices and is often difficult to quantify. However, their influence on the local flow properties is large as flow velocities scale with the square of the channel height. Embedding a color-changing gel into microfluidic channels could help avoid these effects.

\subsection*{Dynamic response}

The dynamic response of the device \RE{under flow} was tested by imposing an oscillating pressure at the channel inlet for the chip shown in Fig. \ref{fig:2}. The imposed pressure profile and the response of the first large sensor (closest to the inlet) are shown in the Supplementary Materials, together with the experimental details. 
\RE{The sensor exhibits an oscillating response with a time delay of about \SI{1.3}{\s} evaluated as the typical time scale for the exponential increase in measured pressure after a step increase in applied pressure. A time shiftof around 1s between the measured and applyed pressure oscillations is observed and found to be of comparable order of magnitude.}. This response time is quite large, in particular compared to that of the gel (\SI{0.1}{\ms}\cite{Yue2014mechano}). We emphasize that this is not the response time of the sensing element itself (which is difficult to access on our device), but rather that of the whole chip. Thus, it is sensitive to all compliant parts in the chip and is simply an upper bound for the response time of the sensing element. This response time could be optimized by working with stiffer materials or by minimizing the volume contained in the sensing cavities. We observe that the shape of the oscillations is conserved, indicating that there is no visible hysteresis in the sensor response for this pressure range and experimental time scale.

\section*{Discussion and conclusion}

We have presented a novel type of pressure sensor, directly using the color change in the visible spectra of a soft hydrogel. This gel can be integrated into sensing units, where a pressure-sensing cavity is separated from the hydrogel by a thin membrane. These multilayered sensing units can be fully integrated into PDMS microchips and allow remote sensing of pressure at given positions in a flow channel. Once calibrated, the read-out is very easy and precise and can be done via a simple color camera. The hydrogel being very soft, we have demonstrated the possibility of measuring pressure in the range between \SI{20}{\milli\bar}-\RE{\SI{140}{\milli\bar }} with a resolution of \SI{10}{\milli\bar} \RE{using two different sensor sizes.} The pressure range can be easily extended to larger pressures by playing with the coupling between the \RE{bending} stiffness of the membrane and the elastic gel. For large cavity sizes the \RE{role of the membrane gets more and more negligible and the} gel deformation under pressure is solely given by the gel \RE{Young's} modulus. Extending the range of pressures accessible to smaller pressures thus either requires working with even softer gels or improving the resolution of the color change readout. \RE{The use of a spectrometer might improve this resolution and avoid the limitations resulting from the saturation of the hue values occurring at strong gel compression and thus towards violet colors.}   

\RE{A more detailed modeling of the elastic coupling of the membrane gel system for different sensor geometries might in the futur allow to achiev even more precise control of the sensor resolution and working range. The full understanding of the time response of the sensor requires additional investigation addressing in particular the interaction between the elastic chip/sensor system and the viscous fluid flow. Such an understanding is required to work with non stationnary or turbulent flows, but is outside the scope of the present work.}

\RE{Our sensor distinguishes itself by its compatibility with diverse fluid types (liquids and gases) as they do not enter into contact with the sensing gel. The inherent design of the sensor ensures the absence of liquid/gas interfaces that could induce measurement errors for small absolute pressures due to the capillary pressure jump.} The design is however not limited to the use of PDMS as presented here, but could easily be fabricated using glass or another type of polymeric materials. The membrane separating the sensing unit from the hydrogel could also be replaced by a different material, even a thin metallic sheet could be imagined as long as the pressure range to be evaluated is sufficient to deform the coupled membrane and gel system. For non-transparent membranes, the color-changing gel has to be observed from below the chamber containing the gel and not as done here from above. 

The use of our pressure sensor is not only limited to microfluidic applications, but it can be plugged into any fluid circuit. For less precise applications a read-out by eye can be imagined and could be useful for example by giving a warning signal above a certain pressure threshold. The spatially resolved pressure measurement opens numerous possibilities for fundamental or industrial research as local pressure measurements in turbulent or complex fluid flows, droplet generation, micro-reactors, cellular growth and deformation, organ-on-chip technology, or medical applications, but also validation of numerical models.

\section*{Methods}

\subsection*{Preparation and characterization of the soft photonic gels}

\subsubsection*{Gel preparation}

The pressure-sensing gel is fabricated similarly to a previously reported lamellar sheet hydrogel \cite{Haque2010}.  It consists of uniaxially aligned reflective bilayers of PDGI (homo-polymerized poly(dodecyl glyceryl itaconate)) and a chemically crosslinked elastic hydrogel matrix of PAAm (polyacrylamide). A large sheet of soft PDGI/PAAm hydrogel with this lamellar bilayer structure parallel to the sheet surface is fabricated by slightly modifying the procedure described by Haque \textit{et al.} \cite{Haque2010}. Briefly, a polymerization chamber is made of two parallel glass plates separated by \SI{0.1}{\mm} thick silicone spacers. The reaction chamber possesses an inlet hole to be filled with the precursor solution (aqueous mixture of \SI{0.1}{\Molar} DGI, \SI{0.0025}{\milli\Molar} SDS, \SI{2.0}{\Molar} AAm, \SI{2}{\milli\Molar} crosslinker and \SI{2}{\milli\Molar} photo-initiator) and an outlet hole connected to an automatic suction pump by a polyethylene tube. Prior to the polymerization, the precursor solution is sucked out of the chamber causing strong shear in the narrow gap between the two glass plates and thousands of lamellar bilayers of self-assembled DGI are formed, perfectly aligned in the direction parallel to the surface of the glass substrate. In order to get stable PDGI bilayers in the hydrogel, this step is followed by \RE{immediate} polymerization by exposing the reaction chamber to UV light (\SI{365}{\nm}) at \SI{50}{\celsius} for \SI{8}{\hour} under an inert Ar atmosphere. \RE{N,N-Methylenebis(acrylamide) (MBAA) was used as crosslinker and Irgacure as initiator.} After polymerization, PDGI bilayers are trapped inside the PAAm matrix. The PDGI/PAAm gels synthesized in this way are then swollen in bi-distilled water for one week with regular freshwater replacement to avoid residuals. To further reduce the moduli of the equilibrium swollen PDGI/PAAm in a second step, the parent PDGI/PAAm gels of \SI{100}{\um} thick were hydrolyzed using a modified procedure that was described previously by Yue \textit{et al.} \cite{Yue2014mechano}. The parent gels were soaked in \SI{1}{\Molar} KOH(aq) aqueous solution for \SI{30}{\min}, followed by heating in an incubator at \SI{50}{\celsius} for \SI{5}{\min}. The process of hydrolysis partially changes the amide functional groups (PAAm) of the parent PAAm hydrogel layers to sodium carboxylate groups (PAAc-Na). The obtained hydrogel, which contains partially hydrolyzed PAAm (PAAcNa), was then extensively washed several times with deionized water to achieve an equilibrium swelling state. The strong swelling taking preferably place along the direction of the layers splits the hard bilayers into smaller domains. Subsequently, the rigid layers continue to reflect visible light, but play no role in the gel deformation. The softness of this hydrolyzed PDGI/PAAm gel is controlled by maintaining the pH of the water. To be used as microfluidic pressure sensors the soft photonic hydrogel of thickness \SI{140}{\um} was maintained in a buffer solution at pH $= 6.92$ using \SI{0.2}{\Molar} KOH(aq) and \SI{0.2}{\Molar} KH2PHO4(aq).

\subsubsection*{Gel characterization} 

At rest, the swollen hydrolyzed gels have an \RE{orange} color. Under compression, the distance between the light-reflecting bilayers is reduced leading to a change in color from \RE{orange} to blue, covering the whole range of visible light. The relation between the applied pressure and the consequent color change, as well as the moduli of the fabricated gels, were characterized using a custom-made setup combing commercial mechanical testing equipment (Tensilon RTC-1310A, Orientec Co.) together with movable angle reflection measurement optics (Hamamatsu Photonics KK, C10027A10687) coupled to a photonic multichannel analyzer (Hamamatsu Photonics KK, C10027) \RE{(see Supplementary Fig. 1~(a)).} Disc-shaped samples of diameter \SI{10}{\mm} were cut out using a gel cutter (model JIS-K6251-7). Samples were placed on a black substrate and compressive loading was applied using the tensile machine. White light was used to illuminate the gel throughout the experiments. The reflection spectrum under compressive loading was acquired by keeping both the angles of incident and reflection at \SI{20}{\degree}.  
The wavelength at maximum reflection intensity was obtained from the reflection spectrum \RE{(see Supplementary Fig. 1~(b) for an example).} The applied compression force, $F$  was registered for each step, together with the real area $A$ of the compressed disc, allowing to calculate the applied pressure $P=F/A$. In this way, the color change as a function of applied pressure was obtained (see Fig. \ref{fig:1} and Supplementary Fig.~2(a)). During compression, the change in thickness of the gel was also registered from the tensile machine. Note that the thickness change can also be deduced from the color change, corresponding to a well-defined change in spacing between the reflecting bilayers. Both methods lead to similar values for the gel compression \RE{\cite{Yue2014mechano, Haque2010}}. From these stress versus strain curves the \RE{Young's} moduli can be measured and have been found to range between \SI{10}{\kilo\pascal} and \SI{100}{\kilo\pascal} \RE{(see Supplemantary Fig. 2~(b))}. \RE{These values are slightly higher compared to the \SI{3.2}{\kilo\pascal} given in \cite{Yue2014mechano} and result from slight changes in the hydrolysis time, swelling and buffer conditions. Gels with slightly higher moduli were more suited to be used in the microfluidic application as they were easier to handle during chip assembly.}

\subsubsection*{Device fabrication}

The device is composed of two main PDMS parts: the top part corresponding to the fluid channel and cavities and the bottom part to the gel \RE{chamber}. Both are obtained by pouring a Sylgard 184 PDMS mixture on \RE{silicon wafers using photolithography and pre-designed molds following the standard procedure. The lithography parameters are adjusted to the desired final height and later confirmed through profilometer measurements. The error in the gel thickness as well as the wafer fabrication have to be taken into account and adjusted in a way that the PDMS gel chamber height remains minimum equal to or slightly higher than the gel thickness to prevent pre-compression. A higher chamber leads to an offset of the pressure. The poured PDMS} is vacuumed to remove bubbles and cured at \SI{70}{\celsius} for \SI{2}{\hour}. \RE{The two parts are separated by a commercially available thin PDMS film (Silicon Sheet, GFSC 6000-100 µ -300Q) purchased from “TOMITA MATEQS, Japan”. } 

To assemble the device, we first put the top PDMS part containing the cavities and a thin PDMS membrane of \SI{100}{\um}, in the \RE{oxygen} plasma cleaner for \SI{5}{\min}, \RE{at \SI{0.2}{\milli\bar}, \SI{20}{W} and \SI{50}{\kilo\Hz}.} \RE{The pieces are} assembled together and cured at \SI{90}{\celsius} for \SI{5}{\min}. 
This operation is repeated with the assembled top \RE{cavities}/thin membrane and bottom PDMS part containing the gel \RE{chambers}. When removed from the plasma cleaner, just before assembling, the pieces of gel are cut with a blade and put in the gel \RE{chamber}.
At last, the final parts are aligned, assembled, and cured at \SI{90}{\celsius} for \SI{5}{\min}. Finally we add buffer in the gel \RE{chamber} to make sure it is well hydrated and we clog the holes to avoid evaporation. \RE{The gel chamber was always kept filled with a buffer with closed outlets and refilled if the buffer partially evaporated.}

\subsubsection*{Microfluidic setup}

The pressure-sensing unit was combined with microfluidic devices. It was observed under a binocular microscope with a white light source oriented perpendicularly toward the gel sample. The observation was made from "above" through the pressure-sensing unit. A black background was insured with black isolation tape at the bottom of the device (below the gel \RE{chamber}) so that the reflected light gave a better color contrast. The images were acquired using a reflex Nikon camera, mounted on the binocular microscope. A pressure controller (Fluigent Lineup Flow EZ) was used to impose controlled static pressure for the sensor calibration, in a range from \SI{0}{\milli\bar} up to \RE{\SI{140}{\milli\bar}}. For flow experiments, a steady flow was imposed inside the microfluidic channel with a syringe pump (Cetoni) at a controlled flow rate.

\bibliography{BiblioSensorAnke}

\section*{Acknowledgements}

We thank Benoit Roman for the useful discussions. This work was supported by Global Station for Soft Matter, Global Institution for Collaborative Research and Education (GSS, GI-CoRE) Hokkaido University, and JSPS KAKENHI (Grant no. 22H04968, 22K21342). We acknowledge support from the ERC Consolidator Grant PaDyFlow under grant agreement 682367. We thank Institut Pierre-Gilles de Gennes (Investissements d'avenir ANR-10-EQPX-34) and AAP Carnot 2021 et PSL Valorisation - Qlife 2022. This research was partially supported by the UGC Research Grant (2023-2024), University of Dhaka, Dhaka-1000, Bangladesh.

\section*{Author contributions statement}

AH, LD, JPG, and AL designed the study. AH and MI fabricated the color-changing gel. AH, LD, and MG performed experiments and analyzed the results. LD, AH, MG, and AL wrote the manuscript. All authors reviewed the manuscript. 

\section*{Additional information}

The authors declare no competing interests. 

\section*{Data availability}

All data generated or analyzed during this study are included in the supplementary information files.

\end{document}